\documentclass{article}

\usepackage{PRIMEarxiv}

\usepackage[utf8]{inputenc} 
\usepackage[T1]{fontenc}    
\usepackage{hyperref}       
\usepackage{url}            
\usepackage{booktabs}       
\usepackage{amsfonts}       
\usepackage{nicefrac}       
\usepackage{microtype}      
\usepackage{lipsum}
\usepackage{fancyhdr}       
\usepackage{graphicx}       
\graphicspath{{media/}}     

\title{Key Expansion Based on Internet X.509 Public Key Infrastructure for Anonymous Voting}

\author{
  Abel C. H. Chen \\
  Chunghwa Telecom Laboratories\\
  \texttt{chchen.scholar@gmail.com} \\
}

\begin{document}
\maketitle

\begin{abstract}
This document focuses on developing a key expansion method based on the internet X.509 public key infrastructure and elliptic curve cryptography, which is applied in the context of anonymous voting. The method enables end entities to maintain anonymity from other end entities, the registration authority, and the certificate authority, while still allowing the validity of end entity certificates to be verified, thereby facilitating anonymous voting services.
\end{abstract}


\section{Introduction}
Given that privacy protection is a critical issue in Internet services, safeguarding user privacy during the process of accessing such services can enhance user satisfaction and engagement. In light of this, this document develops a key expansion method built upon the existing internet X.509 public key infrastructure (PKI), with its security assured through elliptic curve cryptography (ECC) and its ability to ensure anonymity via key expansion. Therefore, this section will introduce the internet X.509 public key infrastructure and elliptic curve cryptography, followed by sequent sections covering conventions and definitions, key expansion, anonymous voting, and case studies.

\subsection{Internet X.509 Public Key Infrastructure}
This document references the internet X.509 public key infrastructure as defined in \cite{RFC5280}, encompassing end entities (EEs), registration authorities (RAs), and certificate authorities (CAs). The end entity generates a key pair, consisting of a private key and a public key, and sends a certificate request to the registration authority. This certificate request includes the end entity's public key. The registration authority reviews the end entity's eligibility to access the application service and, upon verification, forwards the certificate request to the certificate authority. The certificate authority then verifies the correctness of the request and issues the end entity’s certificate, which includes the public key of the end entity.

\subsection{Elliptic Curve Cryptography}
This document references the elliptic curves as defined \cite{RFC5639} and \cite{RFC5480}, which include a prime number $p$, a coefficient $A$, a coefficient $B$, the base point $G = (x_G,y_G)$, the prime order  $n$, and a cofactor $h$. Specifically, let $p > 3$ be a prime and $GF(p)$ a finite field with $p$ elements. The set of solutions $(x,y)$ satisfies the equation $E: y ^ 2 = x ^ 3 + A \times x + B \pmod p$. In asymmetric cryptography applications, a random positive integer $i$ can be generated, where $i$ is within the range $[1, n - 1]$, serving as the private key. The public key $I$ is then derived from the private key $i$ and the base point $G$ as $I = i \times G$.

For instance, with Curve-ID: brainpoolP256r1, by randomly generating the following positive integer $i$ as the private key, the corresponding parameter values and the elliptic curve coordinates of the corresponding public key $I$ are as follows. This key pair can be applied in the signature algorithms (e.g. Elliptic Curve Digital Signature Algorithm (ECDSA)) and the encryption algorithms (e.g. Elliptic Curve Integrated Encryption Scheme (ECIES)).

$p = A9FB57DBA1EEA9BC3E660A909D838D726E3BF623D52620282013481D1F6E5377$

$A = 7D5A0975FC2C3057EEF67530417AFFE7FB8055C126DC5C6CE94A4B44F330B5D9$

$B = 26DC5C6CE94A4B44F330B5D9BBD77CBF958416295CF7E1CE6BCCDC18FF8C07B6$

$G = (x_G,y_G) = \\
  (8BD2AEB9CB7E57CB2C4B482FFC81B7AFB9DE27E1E3BD23C23A4453BD9ACE3262, \\
   547EF835C3DAC4FD97F8461A14611DC9C27745132DED8E545C1D54C72F046997)$

$n = A9FB57DBA1EEA9BC3E660A909D838D718C397AA3B561A6F7901E0E82974856A7$

$h = 1$

$i = 5D98BD1F7985FC560A8963D6709AAC8B01017D02FB14B12CEF168C9662056874$

$I = (x_I,y_I) = \\
  (719CE2A5F8D8174418C3B3AA2E9C4F0EE8AF17F3A9E02A0656E03C32EC05383A, \\
   2E98CA241C58A4933AEE7D4A22394D27EFC1C64618686A00519CC5CB4DE1A93D)$

\section{Conventions and Definitions}
The key words "MUST", "MUST NOT", "REQUIRED", "SHALL", "SHALL NOT", "SHOULD", "SHOULD NOT", "RECOMMENDED", "NOT RECOMMENDED", "MAY", and "OPTIONAL" in this document are to be interpreted as described in BCP 14 \cite{RFC2119}, \cite{RFC8174} when, and only when, they appear in all capitals, as shown here.

\section{Key Expansion for Anonymous Voting}
This document references \cite{IEEE1609} to propose a key expansion method for anonymous voting, consisting of four steps: (1) the generation of an original key pair by the end entity, (2) the generation of a temporary public key by the registration authority, (3) the generation of a formal public key by the certificate authority, and (4) the generation of a temporary private key and a formal private key by the end entity. The temporary public key is generated based on the expanded original public key, and the formal public key is generated based on the expanded temporary public key. Similarly, the temporary private key is generated from the expanded original private key, and the formal private key is generated from the expanded temporary private key. The specific details are presented in the following subsections.

\subsection{The Generation of Original Key Pair by End Entity}
First, the end entity generates an original key pair based on elliptic curve cryptography, consisting of an original private key $i$ and an original public key  $I$ (shown as follows). The end entity holds a valid pre-existing X.509 certificate, which includes the original public key $I$, and possesses the corresponding original private key $i$. The process for obtaining the valid pre-existing X.509 certificate is beyond the scope of this document.

$I = i \times G$

Subsequently, the end entity generates a random integer $r_{RA}$ within the range $[1, n - 1]$ and encrypts it using the public key of the registration authority, resulting in the ciphertext $r_{RA}'$. The end entity also generates an Advanced Encryption Standard (AES) secret key s, which is encrypted using the public key of the certificate authority, resulting in the ciphertext $s'$. The encryption algorithm can be based on ECIES.

The end entity then sends a certificate request to the registration authority, signing the request with its original private key $i$. The certificate request includes the valid pre-existing X.509 certificate, the ciphertext $r_{RA}'$, the ciphertext $s'$, the signature, and other relevant request details.

\subsection{The Generation of Temporary Public Key by Registration Authority}
Upon receiving the certificate request from the end entity, the registration authority verifies the signature using the original public key $I$ and confirms the eligibility of the end entity. Once validated, the key expansion process could be performed.

The registration authority first decrypts the ciphertext $r_{RA}'$ using its private key to obtain the plaintext $r_{RA}$. It then performs key expansion on the original public key $I$ to derive the temporary public key $J$, calculated as follows.

$J = I + r_{RA} \times G$

The registration authority forwards the certificate request to the certificate authority, replacing the original public key $I$ with the temporary public key $J$, removing any personal and sensitive information, and signing the request with its private key. The certificate request includes the temporary public key $J$, the ciphertext $s'$, the signature, and other relevant request details. The registration authority also stores the hash value of the certificate request.

\subsection{The Generation of Formal Public Key by Certificate Authority}
When the certificate authority receives the certificate request from the registration authority, it verifies the signature using the registration authority’s public key and confirms the relevant qualifications. Once validated, the key expansion process begins.

The certificate authority generates a random integer $r_{CA}$ within the range $[1, n - 1]$ and performs key expansion on the temporary public key $J$ to derive the formal public key $K$, calculated as follows. The certificate authority then creates the end entity’s X.509 anonymous certificate, storing the formal public key $K$ in the certificate's public key field.

$K = J + r_{CA} \times G$

Next, the certificate authority decrypts the ciphertext $s'$ using its private key to obtain the plaintext $s$. The AES secret key $s$ is then used to encrypt both the X.509 anonymous certificate and the integer $r_{CA}$, producing the ciphertext $z$. The certificate authority sends a certificate response to the registration authority, signing the response with its private key. The certificate response includes the ciphertext $z$, the signature, the hash value of request, and other relevant response details.

\subsection{The Generation of Temporary Private Key and Formal Private Key by End Entity}
When the registration authority receives the certificate response, it can identify which end entity the request corresponds to based on the hash value of the request and then forward the certificate response to the appropriate end entity. Upon receiving the certificate response, the end entity first verifies the signature using the certificate authority’s public key. Once the signature is verified, the process of private key expansion begins.

The end entity decrypts the ciphertext $z$ using the AES secret key $s$, retrieving the X.509 anonymous certificate and the integer $r_{CA}$. It then expands the original private key $i$ to derive the temporary private key $j$ and further expands the temporary private key $j$ to obtain the formal private key $k$, using the following equations. During the voting process, the end entity can use the formal private key $k$ to sign its ballot. Other end entities can retrieve the formal public key $K$ from the X.509 anonymous certificate and use it to verify the ballot's signature.

$j = i + r_{RA} \pmod n$

$k = j + r_{CA} \pmod n$

\section{Case Studies}
To demonstrate the proposed key expansion method for anonymous voting, two case studies are provided. In the first case study, using a social network platform as an example, each user can connect to the platform using their end entity, which is equipped with an X.509 certificate. The social network platform serves as the registration authority, while an impartial third-party certificate authority issues the anonymous certificates. In the second case study, using a citizen digital certificate as an example, each user can connect to a voting system using their end entity with the citizen digital certificate based on an X.509 certificate. The local district office server acts as the registration authority, and the election commission server serves as the certificate authority, issuing anonymous certificates.

\subsection{The Case Study of Social Network Platform}
In this case study, the user’s device on the social network platform acts as the end entity, while the social network platform server functions as the registration authority. Additionally, an impartial third party serves as the certificate authority, responsible for issuing X.509 anonymous certificates to the end entity. Initially, the user’s device already possesses the original private key $i$ and its corresponding X.509 certificate (which includes the original public key $I$).

When the user wishes to participate in anonymous voting on the social network platform, the user’s device sends a request containing its X.509 certificate to the social network platform server. After verifying the user’s eligibility, the server generates a random integer $r_{RA}$ and expands the original public key $I$ (from the X.509 certificate) to derive the temporary public key $J$. The server then forwards the request, along with the temporary public key $J$, to the impartial third party.

The impartial third party generates a random integer $r_{CA}$ and expands the temporary public key $J$ to derive the formal public key $K$. The impartial third party then issues an X.509 anonymous certificate containing the formal public key $K$, encrypts the X.509 anonymous certificate and the random integer $r_{CA}$, and sends them back to the user’s device. It is important to note that this encryption is done using a shared AES key between the user’s device and the impartial third party, ensuring that only these two entities have access to the X.509 anonymous certificate and the random integer $r_{CA}$.

Finally, the user’s device expands the temporary private key $j$ and the formal private key $k$, allowing the user to sign the ballot using the formal private key $k$. The X.509 anonymous certificate is also included with the ballot. The Curve-ID brainpoolP256r1 is selected in this case study, and the test vectors are shown as follows.

$i = 5D98BD1F7985FC560A8963D6709AAC8B01017D02FB14B12CEF168C9662056874 $

$I = (x_I,y_I) = \\
  (719CE2A5F8D8174418C3B3AA2E9C4F0EE8AF17F3A9E02A0656E03C32EC05383A,\\
   2E98CA241C58A4933AEE7D4A22394D27EFC1C64618686A00519CC5CB4DE1A93D)$

$r_{RA} = \\
     F407A78C2CFC8586AC1BA3199F7CBEF34F138894586B5992B61BB8B5A99C5EE7$ 

$j = A7A50CD00493D820783EFC5F7293DE0CC3DB8AF39E1E63C8151436C9745970B4$

$J = (x_J,y_J) = \\
  (6814044C70048578E6B120480CBA0B81186054403CAE4C67F688F4074AEDF39B,\\
   4969EBCD7400997FBEAF31481DBA738253052A2FF119FE178CF596EFA7AAE156)$

$r_{CA} = \\
     474B007D2533DF88376824D4F784129F7BD7B01F1B3C7E69826912D4121A12DF$ 

$k = 44F4B57187D90DEC714116A3CC94633AB379C06F03F93B3A075F3B1AEF2B2CEC $

$K = (x_K,y_K) = \\
  (80C6DE97A41127BAAFBC4F36E4E514514086A3E4B0F86F9729C52A8767616BF3,\\
   5E333A1B7AC00E2C126C48C343A1A314D2853D4FBD559B9453C434C8C1CDE396)$

\subsection{The Case Study of Citizen Digital Certificate}
In this case study, the citizen possesses a citizen digital certificate, which can be installed on their computer, with the computer acting as the end entity. The citizen digital certificate contains the original private key $i$ and its corresponding X.509 certificate (which includes the original public key $I$). The local district office server serves as the registration authority, while the election commission server acts as the certificate authority, responsible for issuing an X.509 anonymous certificate to the end entity.

When the citizen wishes to cast an anonymous vote in a local representative election, their computer sends a request, including the X.509 certificate, to the local district office server. After verifying the citizen’s eligibility, the local district office server generates a random integer $r_RA$ and expands the original public key I (from the X.509 certificate) to derive the temporary public key $J$. The request, along with the temporary public key $J$, is then forwarded to the election commission server.

The election commission server generates a random integer $r_CA$ and expands the temporary public key $J$ to produce the formal public key $K$. The election commission server then issues an X.509 anonymous certificate containing the formal public key $K$ and encrypts both the X.509 anonymous certificate and the random integer $r_CA$, returning them to the citizen’s computer. It is important to note that during this process, the X.509 anonymous certificate and the random integer $r_CA$ are encrypted using a shared AES key between the citizen’s computer and the election commission server, ensuring that only these two entities have access to this information.

Finally, the citizen’s computer expands the temporary private key $j$ and the formal private key $k$ from the citizen digital certificate. The formal private key $k$ can then be used to sign the ballot, which will also include the X.509 anonymous certificate. The Curve-ID secp256r1 is selected in this case study, and the test vectors are shown as follows.

$i = 5D98BD1F7985FC560A8963D6709AAC8B01017D02FB14B12CEF168C9662056874 $

$I = (x_I,y_I) = \\
  (490D13266EB3E12C28E44C345B345C431D9BFAB5B101D5E0144AB6ECF194D852,\\
   ABC2D769EF6DE7373AD78082BEC46E455A84BAD6CDCDA8FC557438E9AB56ECDD)$

$r_{RA} = \\
     F649BE0670CD1C8325BF4EE06963C680E25140702DBCBAAA3D59B30D6CF3C727 $

$j = 53E27B26EA5318D83048B2B6D9FE730C266BC2C581B9CD5238B674E0D2960A4A$

$J = (x_J,y_J) = \\
  (78BF0F472CD984F8EA7A756A514118652B88224DD344D5D593E03F2AFBCC6FC6,\\
   A4C43CFFC945A1576B290EC63DB8FB31C74EF44F02963C67EB884E025B1A442E)$

$r_{CA} = \\
     A1754E3C913F9C98B1147DD142E66A3CBE59DD5B53B721C39B0A73BCE4EB1B32 $

$k = F557C9637B92B570E15D30881CE4DD48E4C5A020D570EF15D3C0E89DB781257C$ 

$K = (x_K,y_K) = \\
  (6FDFDCE25F876BD4B23FBFBFC9E49872944F01926989B5A72A1FD84125FEB428,\\
   8A108A7CBA97CF26FB9768A7599473F5F9AA27CD462280F41FCAC13FF1BA42E3)$

\section{Security Considerations}
The security considerations of \cite{RFC5639} and \cite{RFC5480} apply accordingly.

The key expansion methods proposed in this document are based on elliptic curve cryptography, meaning their security primarily relies on the difficulty of solving the discrete logarithm problem. Furthermore, the certificate verification methods employed are based on the internet X.509 public key infrastructure, meaning the security limitations associated with the internet X.509 public key infrastructure also apply to this document.

Based on the aforementioned security measures, only the registration authority can decrypt the ciphertext $r_{RA}'$ using its private key to obtain $r_{RA}$. Likewise, only the certificate authority can decrypt the ciphertext $s'$ using its private key to retrieve $s$, and it is the certificate authority that uses $s$ to encrypt and generate the ciphertext $r_{CA}'$, allowing the end entity to obtain the plaintext $r_{CA}$. Since the registration authority does not know the value of $r_{CA}$, it cannot determine the relationship between the temporary public key $J$ and the formal public key $K$. Similarly, because the certificate authority does not know the value of $r_{RA}$, it cannot establish a link between the original public key $I$ and the temporary public key $J$.

\section*{Acknowledgments}
I am pleased to share that this document has been made available as a draft on the IETF website (\url{https://datatracker.ietf.org/doc/draft-chen-x509-anonymous-voting/}). I would like to express my sincere gratitude to the IETF for their consideration of this draft.

\bibliographystyle{unsrt}

\end{document}